\documentclass[preprint,12pt,authoryear,review]{elsarticle}

\usepackage{amssymb,amsmath,amsthm}
\usepackage{algorithm,algpseudocode}
\usepackage{comment}
\usepackage{setspace}
\usepackage{booktabs}
\usepackage{bm}
\usepackage[dvipsnames]{xcolor}
\journal{Journal of Economic Dynamics and Control}
\date{August 31, 2023}

\begin{document}

\begin{frontmatter}

\title{Life cycle insurance, bequest motives and annuity loads}

\author[aleksandar,pavel]{Aleksandar Arandjelovi{\'c}} 
\affiliation[aleksandar]{organization={Institute of Statistics and Mathematical Methods in Economics, TU Wien},
            country={Austria}}

\author[geoff]{Geoffrey Kingston} 
\affiliation[geoff]{organization={Department of Economics, Macquarie University},
            country={Australia}}

\author[pavel]{Pavel V. Shevchenko\corref{cor1}} 
\affiliation[pavel]{organization={Department of Actuarial Studies and Business Analytics, Macquarie University},
            country={Australia}}
\cortext[cor1]{Corresponding author: pavel.shevchenko@mq.edu.au}

\singlespacing

\begin{abstract}
We investigate insurance purchases when bequest motives are age-varying and life insurance and life annuities both carry loads.
The existing life cycle literature assumes bequests are normal goods without being either necessities or luxuries.
Much of the literature also assumes implicitly that life annuity loads are negative.
A key finding of the literature is that the demand for life insurance and the demand for life annuities are symmetrical.
It is optimal to buy life-contingent insurance throughout life, even under loads.
A life annuity phase backs directly onto a life insurance phase.
We find that realistic examples with positive loads on both products reveal up to two distinct periods of non-participation, one in midlife and the other adjoining the maximum age.
We highlight examples with necessity bequests during child-rearing years and luxury bequests thereafter.
This set of assumptions explains why a substantial demand for life insurance during child-rearing years can co-exist with negligible demand for life annuities later on. 
A realistic 18$\%$ load on both products generates this outcome.
\end{abstract}

\end{frontmatter}


\clearpage

\section{Introduction}\label{sec:introduction}
In this paper we investigate numerically the optimal purchases of insurance products in life cycle settings when bequest motives vary with age and life insurance and life annuities both carry loads.
Our main research question is the ``deep" annuity puzzle whereby a thin market for life annuities co-exists with a thick market for life insurance.
The regular version of this puzzle restricts attention to the fact of a thin annuities market.

The life cycle model integrates an individual's life insurance phase with her life annuity phase, recognizing that buying a life annuity is like selling life insurance.
This symmetry is evident when all life-contingent insurance is of the instantaneous-term variety, as is generally the case in the literature under discussion.
The annuity buyer's estate makes a payment to the seller if and when the buyer dies.
Likewise, the seller of life insurance makes a payment to the buyer's estate, contingent on the buyer's death.
\citet{yaari65}, \citet{hakansson69} and \citet{merton71} led the way.
Bequests were assumed to be normal goods without being either necessities or luxuries.
In particular, constant relative risk aversion (i.e. power utility) was assumed whenever bequest utility had a specific functional form.
Insurance was initially assumed to be actuarially fair.

Many subsequent contributions do incorporate loads.
They generally specify loads in terms of markups over actuarially fair insurance.
For this purpose, life cycle models typically introduce a single load parameter that can be interpreted as the ask price of life insurance.
The parameter does double duty, for the life annuity phase as well as the life insurance phase.
Early examples include the influential contributions of \citet{fischer73} and \citet{richard75}.
However, it turns out that the assumptions of a single-parameter model of loads and normal bequests both act to exaggerate the symmetry between the demand for life insurance and the demand for life annuities. 

In the case of loads, we show below that the single-parameter load model of these early contributions (and many subsequent ones) entails negative annuity loads.
As a consequence, the demand for annuities is artificially inflated.
This is unhelpful for resolving the annuity puzzle, deep or otherwise.
Moreover, it turns out that bid-ask spreads in the market for life-contingent insurance are effectively tied to zero.
This implicit assumption leads to a counterfactual prediction that people will participate continuously in the market for life-contingent insurance.

Loads on life-contingent insurance products can be seen as a form of illiquidity or transaction cost.
Such frictions have been studied in the context of risky securities.
See for example \citet{magill76}, \citet{davis90} and \citet{shreve94}.
These contributions analyze positive bid-ask spreads, in contrast to the existing literature on life-contingent insurance over the life cycle.

We introduce a second load parameter that can be interpreted as the bid price of life insurance.
It ensures non-negative annuity loads and enables positive bid-ask spreads.
Moreover, we assume bequests are necessities during child-rearing years.
This is a natural assumption for people with dependents, notably, children at home.
By contrast, bequests are luxuries during retirement years, consistent with a substantial recent body of research. 
Our main contribution, then, is to show numerically that a two-parameter load model and a specification of bequest motives that allows for age variation can jointly resolve the deep annuity puzzle.
Previous contributions have not sought to account for the disparity between the demand for life insurance and the demand for life annuities.

Much of the life cycle literature is primarily concerned with life insurance.
However, the dynamic programming principle suggests that parameters determining the demand for life annuities could feed forward onto the demand for life insurance.
For completeness, we investigate this possibility.
As it turns out, our computations suggest that changes in annuity loads have no effect on the demand for life insurance.
Likewise, changes in life insurance loads have no effect on the demand for annuities.

The paper is organized as follows.
Section \ref{model_sec} sets out an optimization problem.
Section \ref{literaturecomparison_sec} compares and contrasts our approach with previous contributions to the 21st century literature.
Section \ref{sec:piratio} presents a methodology that obtains reasonable estimates for loads.
Section \ref{sec:after} reports results of our computations for the post-retirement phase and Section \ref{sec:fullcycle} reports results for the full life cycle.
Section \ref{sec:conclusions} summarizes our conclusions.
\ref{ap:PY} reports results for the case of a truncated life cycle and \ref{ap:mortrate} calibrates a Gompertz mortality model to data.
\ref{ap:aead} gives details of the age-earnings profile and the age-varying bequest parameter assumed here.
\ref{ap:DP} explains the dynamic programming approach used for the numerical solution of the optimization problem.

\section{Model}\label{model_sec}
Consider a market for life-contingent insurance.
The individual has the opportunity to purchase and sell term insurance on her own life, where the sale of insurance corresponds to the purchase of a life annuity.
Life-contingent insurance is offered continuously, and the individual enters a contract by paying the premium rate $p(t)$, which buys life insurance in the amount $p(t)/\eta(t)$ or $p(t) / \theta(t)$ dollars for the next instant, depending on whether $p(t) > 0$ or $p(t) < 0$.
The mortality rate, $\lambda(t)$, represents the instantaneous death rate for the individual surviving up to time $t$.
Hence insurance is actuarially fair when $\theta(t) = \lambda(t) = \eta(t)$ for all $t$, and it is loaded when $\theta(t) \le \lambda(t) \le \eta(t)$, with at least one of those inequalities being strict for some time $t$.
Under our two-parameter model of insurance loads, then, an individual with financial wealth $W(t)$ who dies at age $t$ leaves a legacy $Z(t)$ given by
\begin{equation}\label{eq:legacy}
Z(t) = \begin{cases} W(t) + \frac{p(t)}{\eta(t)}, & \mbox{if } p(t) > 0, \\ W(t), & \mbox{if } p(t) = 0, \\ W(t) + \frac{p(t)}{\theta(t)}, & \mbox{if } p(t) < 0, \end{cases}
\end{equation}
according as whether she is long, not invested or short in life insurance, the last case corresponding to being long in life annuities.

Bequest utility $B$ is of hyperbolic absolute risk aversion (HARA) form with age-varying shift parameter $c_{b}(t)$,
\begin{equation}\label{eq:bequestutility}
B(t, z) = \Bigl(\frac{\phi}{1-\phi}\Bigr)^{\sigma} \frac{\bigl( \frac{\phi}{1-\phi}c_{b}(t) + z \bigr)^{1-\sigma}}{1-\sigma},
\end{equation}
where $\phi \in (0,1)$ is the marginal propensity to bequeath, and $\sigma>0$ is the coefficient of relative risk aversion for consumption utility.
The parameter $c_{b}(t)$ varies with age as described in Figure \ref{fig:profiles} below, where $c_{b}(t)$ is described as the age-dependency profile and is shown alongside $y(t)$, this paper’s age-earnings profile (or income function), which is standard; see e.g. \citet{mincer74}.
Following \citet{lockwood12,lockwood18} and others, the age-earnings profile incorporates an annuity component, reflecting a Social Security entitlement.
Equations specifying the age-dependency profile and age-earnings profile are set out in \ref{ap:aead}.

\begin{figure}[htb]
\begin{center}
\includegraphics{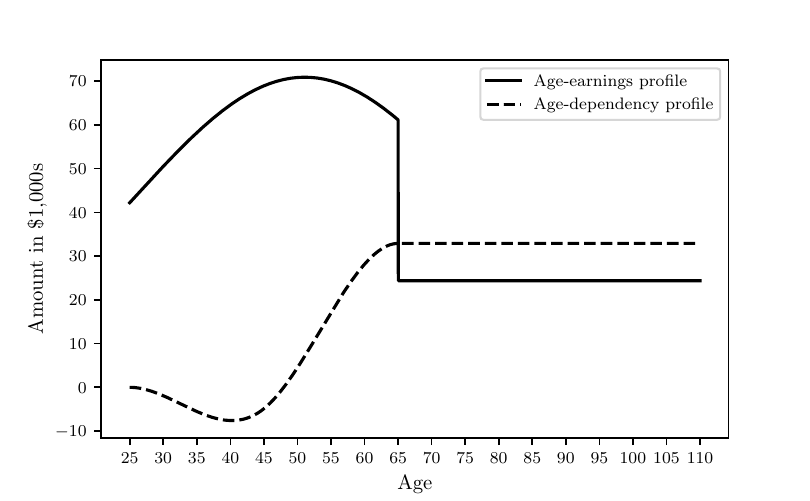}
\vspace*{-1.00cm}
\caption{Age-earnings profile and age-dependency profile}
\label{fig:profiles}
\end{center}
\end{figure}

Recent research finds that power utility is not a good model of the bequest utility of elderly individuals, especially affluent ones.
Rather, bequests by the elderly tend to be luxury goods \citep{carroll98,denardi10,lockwood12,lockwood18}.
These contributions model the bequest motives of the elderly by means of a fixed positive value of $c_{b}(t)$ in Equation \eqref{eq:bequestutility}.
We extend their parametric model of bequest motives, into the individual’s work years, which typically overlap substantially with child-rearing years.
The additive shift parameter has a negative sign during that time. 
In more detail, bequests are necessities rather than luxuries during most of the individual’s working life, becoming more essential in the event the family grows further, but eventually transiting to luxuries as dependents progressively attain independence.
For example, Figure \ref{fig:profiles} postulates that $c_b(t)$ rises smoothly in value from about -\$7,600 at around age 40 (the assumed age of maximum dependency within the household) to \$32,900 at age 65.
Following \citet{lockwood18}, we assume $c_b(t)$ is fixed thereafter.

Since human capital has no closed-form solution here, necessities and luxuries must be characterized by the sign of the additive shift parameter in the bequest utility function rather than the proportionate response of planned bequests to a given increase in the sum of financial wealth and human capital.
This alternative characterization is standard in the literature on luxury bequests by the elderly.
It readily extends to the pre-retirement phase.
For comparison, we also report numeric results for the traditional life-cycle case of power bequest utility. 

Assume that the individual's remaining lifetime, $\tau$, is a random variable with known probability density function $f$.
We introduce a finite planning horizon $T>0$.
The individual's maximum remaining lifespan need not coincide with $T$.
For example, $T$ could denote the time to retirement, as in \citet{pliska07}.
The survival function $\bar{F}(t)$ modeling the probability of survival up to time $t$ is $\bar{F}(t) = \exp(-\int_{0}^{t}\lambda(s)\, \mathrm{d}s)$.
We assume a Gompertz model for $\lambda(t)$.
For a functional form of $\lambda(t)$, see Table \ref{tab:base_params}.
The model was calibrated to mortality rates from the G$12$ countries; details are available in \ref{ap:mortrate}.

The individual consumes at rate $c(t)$.
Instantaneous utility from consumption is of constant relative risk aversion (CRRA) form,
\begin{equation*}
U(c) = \frac{c^{1-\sigma}-1}{1-\sigma}. 
\end{equation*}

Given a pair $(c,p)$ of consumption and insurance plans, the individual's financial wealth $W(t)$ evolves according to 
\begin{equation}\label{eq:wealth}
W(t) = w + \int_{0}^{t} \bigl( rW(s) + y(s) - c(s) - p(s) \bigr)\, \mathrm{d}s,
\quad t \in [0, \tau \wedge T],
\end{equation}
where $w$ denotes the initial level of financial wealth, $r$ denotes the risk-free rate, and $\tau \wedge T = \min\{\tau, T\}$.
In case of death at time $t \le T$, the total legacy $Z(t)$ is given by Equation \eqref{eq:legacy}.
If the individual survives beyond the planning horizon $T$, the legacy equals the financial wealth $W(t)$.
As in previous contributions, we assume that the individual cannot leave a negative legacy.
Given an initial level of financial wealth $w$, the individual's problem is to choose a consumption and insurance plan $(c^\ast,p^\ast)$ that maximizes the expected utility of discounted consumption and bequest,
\begin{equation}\label{eq:optim}
\begin{aligned}
\sup_{(c,p)} \mathbb{E}\Bigl[ \int_{0}^{\tau \wedge T} & \mathrm{e}^{-\beta t}U\bigl(c(t)\bigr)\, \mathrm{d}t \\ & + \mathrm{e}^{-\beta\tau}B\bigl(\tau, Z(\tau)\bigr)\mathrm{1}_{\{\tau \le T\}} + \mathrm{e}^{-\beta T}B\bigl(T, W(T)\bigr)\mathrm{1}_{\{\tau > T\}} \Bigr],
\end{aligned}
\end{equation}
where $\beta$ denotes the rate of time preference, $\mathbb{E}[\,\cdot\,]$ denotes the expectation with respect to the randomness generated by the uncertain lifetime $\tau$, and $\mathrm{1}_{\{\cdot\}}$ denotes the indicator function.
This is a problem that can be solved via dynamic programming; see \ref{ap:DP}.
The model parameters are summarized in Table \ref{tab:base_params} below.

\begin{table}[htb]
\caption{Model parameters}
\begin{center}
\begin{tabular}{@{}ll@{}}
\toprule
\textbf{Model parameter} & \textbf{Value}\\
\midrule
Initial age of the individual & $x=25$ years\\
Risk-free rate & $r=3.2\%$\\
Risk-aversion coefficient & $\sigma=2$\\
Rate of time preference & $\beta=-\ln(0.975)$\\
Propensity to bequeath & $\phi=0.95$\\
Mortality rate & $\lambda(t)=\tfrac{1}{b}\exp\bigl(\tfrac{x+t-m}{b}\bigr)$\\
Modal age at death & $m=88.23$\\
Scale parameter & $b=9.38$\\
\bottomrule
\end{tabular}
\label{tab:base_params}
\end{center}
{\small Notes: Values for the risk-free rate, risk-aversion coefficient, rate of time preference and propensity to bequeath are from \citet{lockwood18}.
Parameters for the mortality rate were calibrated to data from the G$12$ countries; see \ref{ap:mortrate}.
\par}
\end{table}

\section{Comparison with recent literature}\label{literaturecomparison_sec}
\citet{pliska07} make a leading theoretical and numeric contribution to this century's literature on life cycle insurance.
Their theory offers a particularly clear exposition of that literature's single-parameter model of insurance loads. 
We therefore draw upon \citet{pliska07} to explain how the ubiquitous single-parameter model gives rise to an implicit assumption of negative annuity loads and zero bid-ask spreads.
They define $\eta(t)\ge\lambda(t)$ as the ``premium-insurance ratio", that being the single load parameter in question.
As in the preceding section, life-insurance premiums paid at the rate $p(t)$ secure for the upcoming instant a sum insured equal to $p(t)/\eta(t)$.
Note that the parameter in the denominator can be interpreted as the ask price of life insurance.
By assumption, life insurance carries a positive load whenever the inequality $\eta(t)\ge\lambda(t)$ is strict.
So far so good.
However, the single-parameter load model goes on to assume or imply that annuity income at the rate $p(t)$ entails an outlay of $p(t)/\eta(t)$ if the annuitant dies during the upcoming instant, payable by the annuitant’s estate to the provider. 
If the inequality $\eta(t)\ge\lambda(t)$ is strict, this annuity is cheaper than an actuarially fair one.
Annuities are implicitly subsidized.
They make losses for insurance companies and will be in artificially high demand.
For this reason, the preceding section introduced a second premium-insurance ratio, $\theta(t)$, $0<\theta(t)\le\lambda(t)$, for life annuities.
If the individual dies at time $t$, her estate pays the provider an amount $p(t)/\theta(t)$.
The parameter in the denominator can now be interpreted as the bid price of life insurance.
Loads now remain non-negative when the individual goes long in life annuities.
In this way, a two-parameter model of the (additive) bid-ask spread in the insurance market, namely $\eta(t)-\theta(t)$, can ensure that the spread is positive rather than zero.

Zero spreads in the context of a one-parameter model enable what is in effect a complete-markets solution, even when life insurance loads are positive.
There is continuous participation in the market for life-contingent insurance.
The reason is that although life insurance is more expensive than the actuarially-fair benchmark, this is just offset by life annuities being cheaper.
Insurance demands, human capital amounts and the value function all have closed-form solutions.
In these ways, relinquishing zero bid-ask spreads comes at the cost of reduced tractability.
Once the bid-ask spread is positive, we lose effective market completeness and there is no longer continuous participation.
We need to fall back on numerical computations.

\citet{huang08b} consider the case of HARA consumption utility whereby a state-varying shift parameter helps motivate the demand for life insurance. 
Bequests are effectively lifelong necessities.
This life cycle setup is investigated both theoretically and numerically. 
The demand for annuities is found to be positive well before retirement, under either zero loads or loads described by a one-parameter model.
People participate continuously in the market for life-contingent insurance.

\citet{lockwood12,lockwood18} confines attention to the retirement phase.
He addresses the annuity puzzle, combining luxury bequest utility with positive annuity loads.
He estimates the relevant shift parameter in the bequest utility function, and goes on to show numerically that the demand for annuity-type products is weak when there are luxury bequests in conjunction with realistic loads.
 
The calibrations reported by \citet{pashchenko2013accounting} also deal exclusively with the retirement phase.
She reports that only 5\% of the singles aged 70 in her sample own a life annuity.
By the same token, there is considerable dispersion in annuity purchases across income quintiles.
An ``administrative load" of 10\% applies to annuity income.
Loads also arise from adverse selection.
There is a calibrated model of this source of loads on annuities, although the resulting total loads are not spelled out.
Other factors depressing demand are luxury bequests, medical expenses, public annuity-like income, illiquid housing wealth, minimum purchase requirements and preannuitised wealth.
Pashchenko's headline calibrated model predicts that 20\% of singles aged 70 will participate in the market.
In this way, \citet{pashchenko2013accounting} accounts for most of her sample's non-participation in annuity markets.
No single factor dominates the others as a prime cause of weak demand.

The classic overlapping generations model considers social welfare in an economy where bequests are not merely luxuries but provide no utility for the legator.
Put another way, bequest motives are ``inoperative".
Even if the proceeds of unintended bequests are recycled by the government, the relative price of delayed consumption is too high, social saving is suboptimal, and capital formation is too low.
Recent investigations of this scenario include \citet{feigenbaum2013really} and \citet{heijdra2014tragedy}.
Our notion of necessity bequests during child-rearing years suggests that there may be offsetting tendencies at work.
Life insurance promotes saving.\footnote{We thank an anonymous referee for this observation.}
However, we leave this question to future research.

\citet{peijnenburg16} investigate the annuity puzzle in the context of a full (adult) life cycle.
They show numerically that if (i) bequests are lifelong normal goods and (ii) insurance is actuarially fair, then the annuity puzzle persists under various circumstances that had previously been regarded as lessening it.
By way of comparison, we show numerically that if (i) bequests evolve from being necessities to luxuries and (ii) all life-contingent insurance products carry a moderate positive load (e.g. 18\% at age 65), then the annuity puzzle disappears, including the ``deep" version of it. 

\section{Loads}\label{sec:piratio}
Following \citet{brown07} and others, we specify loads as fixed markups over the benefits that would be payable were insurance actuarially fair.
For example, \citet{brown07} find that a typical care insurance policy purchased by a 65 year old carries an 18$\%$ load.
This means that for every dollar the individual pays in premiums, she expects care benefits worth 82 cents in present-value terms.
Loads applying at age 65 together with Gompertz mortality serve to pin down lifelong load schedules.
For this purpose we need to generalize the Brown--Finkelstein approach to incorporate life insurance.
That turns out to be straightforward.
For robustness, we consider a range of loads for both life insurance and, especially, life annuities.
We highlight the case of a single lifelong load schedule pinned down by an 18$\%$ load on both products when bought by individuals aged 65.
This setup enables a parsimonious numeric resolution of the deep annuity puzzle.

\citet{brown07} confine attention to long-term care insurance policies covering home health care, assisted living and nursing home residence.
That we treat their reported 18$\%$ load as our base case for term life annuities is in line with the estimate of \citet{lockwood18}.
A load of 14$\%$ at age 65 is also noteworthy, as it turns out to be just sufficient to extinguish the demand for life annuities.
A range of loads and their implications for the model parameters are shown in Figure \ref{fig:loads} and Table \ref{tab:load_kappas} below.

In this section we present a methodology that allows us to obtain reasonable estimates for $\eta(t)$ and $\theta(t)$, starting from the assumption of an 18$\%$ load on insurance benefits for a 65 year old.
For clarity, we henceforth use the subscripts $\emph{ins}$ and $\emph{ann}$, depending on whether we are studying life insurance or life annuity loads.

\subsection{Life annuity loads}
Consider the expected net present value of a loaded, perpetual life annuity contract with continuous $\$$1 of payments, $\bar{a}(\kappa_{\mathrm{ann}})$.
Here, $\kappa_{\mathrm{ann}} \ge 1$ denotes the mortality loading factor, with $1/\kappa_{\mathrm{ann}}$ being applied as a multiplicative factor to the mortality rate $\lambda(t)$.
The case $\kappa_{\mathrm{ann}}=1$ corresponds to no load, while $\kappa_{\mathrm{ann}} > 1$ decreases mortality risk by reducing the mortality rate, thereby extending longevity and increasing the value of the life annuity contract.

Assuming a Gompertz model for $\lambda(t)$ with parameters from Table \ref{tab:base_params},
\begin{align*}
\bar{a}(\kappa_{\mathrm{ann}}) & = %
\int_{0}^{\infty} \exp\Bigl(-rt-\int_{0}^{t}\tfrac{1}{\kappa_{\mathrm{ann}}}\lambda(s)\, \mathrm{d}s \Bigr)\, \mathrm{d}t \\ & = %
\exp(C) \int_{0}^{\infty}\exp\Bigl(-rt-C\exp\bigl(\tfrac{t}{b}\bigr)\Bigr)\, \mathrm{d}t,
\end{align*}
where $C = \exp((x-m)/b)/\kappa_{\mathrm{ann}}$.
In order to bring the last integral into a more tractable form, we substitute $u = \exp(t/b)$,
\begin{align*}
\bar{a}(\kappa_{\mathrm{ann}}) & = %
b \exp(C) \int_{1}^{\infty}u^{-(1+rb)}\exp(-Cu)\, \mathrm{d}u \\ & = %
b \exp(C) E_{1+rb}(C),
\end{align*}
where $E_{s}(z)$ denotes the generalized integro-exponential function \citep{milgram85}.

Assume that the individual purchases a life annuity contract by paying a single upfront premium $P$.
For every $\$1$ of premium paid, the insurance company will only pay $\$ (1-L_{\mathrm{ann}})$ of benefits, where $L_{\mathrm{ann}} \in (0,1)$ denotes a load.
Given a load $L_{\mathrm{ann}}$, we are thus looking for a mortality loading factor $\kappa_{\mathrm{ann}}$, such that
\begin{equation}\label{eq:solve_ann}
(1-L_{\mathrm{ann}})\, \bar{a}(\kappa_{\mathrm{ann}}) = \bar{a}(1),
\end{equation}
where $\bar{a}(1)$ is the expected net present value of benefits to be paid to the individual, and $P = \bar{a}(\kappa_{\mathrm{ann}})$ is the premium that is charged by the insurance company.

Provided that Equation \eqref{eq:solve_ann} admits a solution, we define our premium-insurance ratio for the purchase of life annuities as $\theta(t) = \lambda(t)/ \kappa_{\mathrm{ann}}$ and observe that, by construction, the inequality $\theta(t) \le \lambda(t)$ is always satisfied.
Moreover, note that
\begin{equation*}
\theta(t) = \frac{1}{b} \exp\Bigl( \frac{x+t-(m+\ln(\kappa_{\mathrm{ann}}^{b}))}{b} \Bigr).
\end{equation*}
In other words, charging the premium-insurance ratio $\theta(t)$ corresponds to increasing the modal age at death from $m$ to $m+\ln(\kappa_{\mathrm{ann}}^{b})$ in our model for the mortality rate $\lambda(t)$.

\subsection{Life insurance loads}
Consider the expected net present value of a loaded, perpetual life insurance contract with a payment of $\$$1 at the moment of death, $\bar{A}(\kappa_{\mathrm{ins}})$.
The mortality loading factor $\kappa_{\mathrm{ins}}$ is now being applied directly as a multiplicative factor to the mortality rate $\lambda(t)$.
The case $\kappa_{\mathrm{ins}}=1$ corresponds to no load, while $\kappa_{\mathrm{ins}} > 1$ increases mortality risk by increasing the mortality rate, thereby shortening longevity and increasing the value of the life insurance contract.

Assuming a Gompertz model for $\lambda(t)$ with parameters from Table \ref{tab:base_params},
\begin{align*}
\bar{A}(\kappa_{\mathrm{ins}}) & = %
\int_{0}^{\infty} \exp\Bigl(-rt -\int_{0}^{t}\lambda(s)\kappa_{\mathrm{ins}}\, \mathrm{d}s \Bigr)\lambda(t)\kappa_{\mathrm{ins}}\, \mathrm{d}t\\%
& = \frac{C}{b} \exp(C) \int_{0}^{\infty}\exp\Bigl(-(r-\tfrac{1}{b})\, t-C\exp\bigl(\tfrac{t}{b}\bigr)\Bigr)\, \mathrm{d}t,
\end{align*}
where $C = \exp((x-m)/b)\kappa_{\mathrm{ins}}$.
In order to bring the last integral into a more tractable form, we again substitute $u = \exp(t/b)$,
\begin{align*}
\bar{A}(\kappa_{\mathrm{ins}}) & = C \exp(C) \int_{1}^{\infty}u^{-rb}\exp(-Cu)\, \mathrm{d}u \\ & = C \exp(C) E_{rb}(C).
\end{align*}

Assume that the individual buys a life insurance contract by paying a single upfront premium $P$.
Given a load $L_{\mathrm{ins}}$, we are thus looking for a mortality loading factor $\kappa_{\mathrm{ins}}$, such that
\begin{equation}\label{eq:solve_ins}
(1-L_{\mathrm{ins}}) \cdot \bar{A}(\kappa_{\mathrm{ins}}) = \bar{A}(1),
\end{equation}
where $\bar{A}(1)$ is the expected net present value of benefits to be paid to the individual, and $P = \bar{A}(\kappa_{\mathrm{ins}})$ is the premium that is charged by the insurance company.

Provided that Equation \eqref{eq:solve_ins} admits a solution, we define our premium-insurance ratio for the purchase of life insurance as $\eta(t) = \lambda(t)\kappa_{\mathrm{ins}}$ and observe that, by construction, the inequality $\lambda(t) \le \eta(t)$ is always satisfied.
Moreover, note that
\begin{equation*}
\eta(t) = \frac{1}{b} \exp\Bigl( \frac{x+t-(m-\ln(\kappa_{\mathrm{ins}}^{b}))}{b} \Bigr).
\end{equation*}
In other words, charging the premium-insurance ration $\eta(t)$ corresponds to decreasing the modal age at death from $m$ to $m-\ln(\kappa_{\mathrm{ins}}^{b})$ in our model for the mortality rate $\lambda(t)$.

\subsection{Calibrating loads}
We aim to find reasonable values of $\eta(t)$ and $\theta(t)$.
\citet{brown07} estimate that the load for a policy purchased by a 65 year old is 18$\%$ if held until death.
For this reason, we fix $x=65$ as well as $L_{\mathrm{ins}} = L_{\mathrm{ann}} = 18\%$.
Furthermore, we set 2$\%$ for the risk-free rate $r$ used for discounting, a value that has also been used by \citet{lockwood18} for calculating insurance premiums and benefits.

We solve Equations \eqref{eq:solve_ann} and \eqref{eq:solve_ins} numerically via a root-finding algorithm, resulting in $\kappa_{\mathrm{ann}} = 2.0377$ for life annuities, corresponding to an increase of the modal age at death to $m_{\mathrm{ann}} = 94.91$, and $\kappa_{\mathrm{ins}} = 4.7446$ for life insurance, corresponding to a decrease of the modal age at death to $m_{\mathrm{ins}} = 73.63$.

\begin{figure}[htb]
\begin{center}
\includegraphics{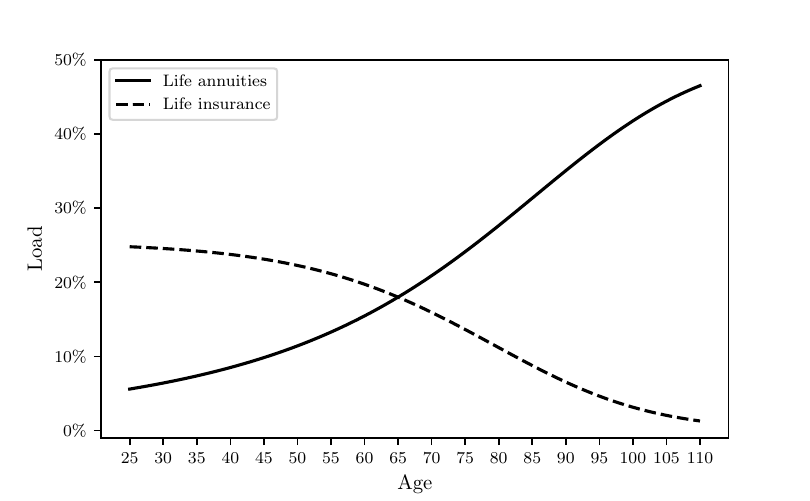}
\vspace*{-1.00cm}
\caption{Loads $L_{\mathrm{ann}}$ and $L_{\mathrm{ins}}$ by age of purchase, with $\kappa_{\mathrm{ann}} = 2.0377$ and $\kappa_{\mathrm{ins}} = 4.7446$}
\label{fig:loads}
\end{center}
\end{figure}

Having fixed the mortality loading factors $\kappa_{\mathrm{ann}}$ and $\kappa_{\mathrm{ins}}$ to the values above, we compute the implied loads for ages other than $x=65$ by solving Equations \eqref{eq:solve_ann} and \eqref{eq:solve_ins} for $L_{\mathrm{ins}}$ and $L_{\mathrm{ann}}$ respectively.
The evolution of those loads over time is presented in Figure \ref{fig:loads}.
Fixed markups over fair insurance together with Gompertz mortality entail loads that vary with the buyer's age, as noted by \citet{brown07}, among others. 
The curve for life annuities qualitatively resembles Figure 1 in \citet{brown07}.
While the differences between the two figures might be explained by differing assumptions about the risk-free rate $r$ (we use $r=2\%$ to match the setting of \citet{lockwood18}, while \citet{brown07} use the term structure of yields of U.S. Treasury strips), we conclude that our model for premium-insurance ratios is broadly consistent with practice.

The choice of load $L$ has a direct impact on the premium-insurance ratios $\eta(t)$ and $\theta(t)$.
Table \ref{tab:load_kappas} sets out mortality loading factors and their implied modal ages at death which were calibrated at the age 65 for different loads $L$.
These are inputs to Table \ref{tab:demand}, which sets out annuity demands corresponding to different loads and ages.

\begin{table}[htb]
\caption{Mortality loading factors $\kappa_{\mathrm{ins}}$, $\kappa_{\mathrm{ann}}$ and corresponding implied modal ages at death $m_{\mathrm{ins}}$, $m_{\mathrm{ann}}$ for different choices of load $L$}
\begin{center}
\begin{tabular}{@{}ccccc@{}}
\toprule
$L$ & $\kappa_{\mathrm{ins}}$ & $m_{\mathrm{ins}}$ & $\kappa_{\mathrm{ann}}$ & $m_{\mathrm{ann}}$\\ 
\midrule
0$\%$ & 1.0000 & 88.23 & 1.0000 & 88.23\\
2$\%$ & 1.1482 & 86.93 & 1.0678 & 88.85\\
4$\%$ & 1.3264 & 85.58 & 1.1434 & 89.49\\
6$\%$ & 1.5426 & 84.16 & 1.2280 & 90.16\\
8$\%$ & 1.8081 & 82.67 & 1.3232 & 90.86\\
10$\%$ & 2.1381 & 81.10 & 1.4306 & 91.59\\
12$\%$ & 2.5547 & 79.43 & 1.5527 & 92.36\\
14$\%$ & 3.0903 & 77.65 & 1.6921 & 93.16\\
16$\%$ & 3.7941 & 75.72 & 1.8523 & 94.01\\
18$\%$ & 4.7446 & 73.63 & 2.0377 & 94.91\\
20$\%$ & 6.0742 & 71.31 & 2.2537 & 95.85\\
\bottomrule
\end{tabular}
\end{center}
\label{tab:load_kappas}
\end{table}

\section{After retirement}\label{sec:after}
Consider financial plans made (or remade) at the age of retirement, assumed here to be 65.
\citet{fischer73} showed that CRRA bequest utility generates substantial demand for annuities, even before retirement.
Annuity loads were (implicitly) negative in that study.

\begin{figure}[htb]
\begin{center}
\includegraphics{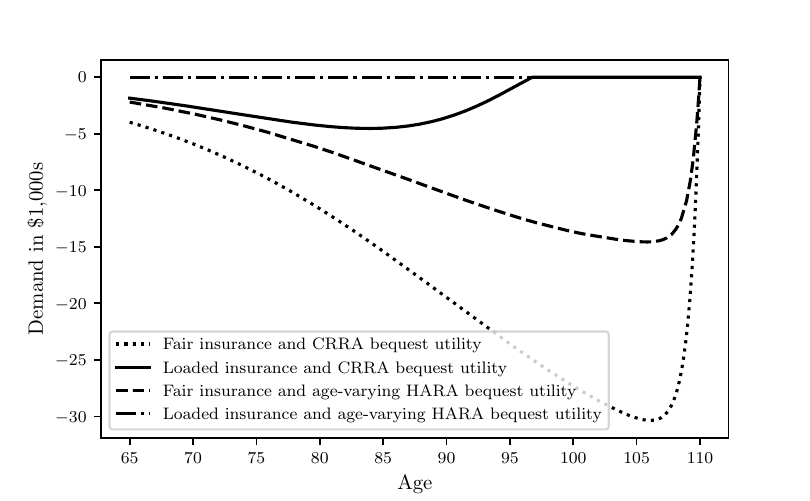}
\vspace*{-1.0cm}
\caption{Optimal life insurance and life annuity purchase after retirement}
\label{fig:postret}
\end{center}
\end{figure}

Figure \ref{fig:postret} shows results of our computations in the case of an 18$\%$ load and financial wealth of $\$$500{,}000 at the time of retirement.
It suggests the following observations.
CRRA bequest utility and fair annuities together generate a strong demand for annuities that begins before retirement (consistent with previous contributions) and lasts for as long as the maximum lifespan of the retiree.
The demand is particularly strong at advanced ages.
Luxury bequest utility and fair annuities together see a considerable fall in the demand for annuities, although demand still begins before retirement and lasts for as long as the maximum lifespan of the retiree.
CRRA bequest utility and an 18$\%$ load on annuities together see not only a further fall in the demand for annuities, but non-participation in the market, beginning at age 97.
Intuition for non-participation is given in the next section.
Finally, luxury bequest utility and an 18$\%$ load see negligible participation in the market for annuities, in line with the findings of \citet{lockwood12,lockwood18}.

A more granular approach reveals the load value that just suffices to extinguish annuity demand in the case of luxury bequests and positive loads.
In particular, Table \ref{tab:demand} shows that a 14$\%$ load just extinguishes annuity demand in this case.
The table also shows that, as we progressively reduce loads below 14$\%$, the demand for annuities at advanced ages rises to substantial levels, even with luxury bequests.
Demand is sensitive to loads.
As a robustness check, we also calibrated annuity demands when wealth at the time of retirement is either reduced to $\$$400{,}000 or increased to $\$$600{,}000.
Broadly speaking, there are commensurate falls and rises in annuity demands.
For example, when the load is 10$\%$ and wealth falls from $\$$500{,}000 to $\$$400{,}000, annuity demand by an individual aged 65 falls from $\$$480 per year to $\$$400 per year.

\begin{table}[htb]
\caption{Load-dependent annuity demand (in $\$100$s) at different ages when financial wealth at the time of retirement is $\$500{,}000$ and bequests are luxuries}
\begin{center}
\begin{tabular}{@{}cccccccc@{}}
\toprule
$L$ & 65 & 70 & 75 & 80 & 85 & 90\\ 
\midrule
0$\%$ & 22.1 & 32.4 & 46.0 & 62.9 & 82.4 & 102.8 \\
2$\%$ & 18.3 & 26.2 & 35.7 & 45.9 & 54.2 & 56.0 \\
4$\%$ & 14.7 & 20.2 & 25.9 & 29.8 & 28.0 & 13.8 \\
6$\%$ & 11.2 & 14.5 & 16.6 & 14.8 & 4.0 & 0.0 \\
8$\%$ & 7.9 & 9.1 & 7.9 & 0.9 & 0.0 & 0.0 \\
10$\%$ & 4.8 & 4.0 & 0.0 & 0.0 & 0.0 & 0.0 \\
12$\%$ & 1.9 & 0.0 & 0.0 & 0.0 & 0.0 & 0.0 \\
14$\%$ & 0.0 & 0.0 & 0.0 & 0.0 & 0.0 & 0.0 \\
\bottomrule
\end{tabular}
\end{center}
\label{tab:demand}
\end{table}

\section{Full life cycle}\label{sec:fullcycle}
Consider financial plans made at the outset of working life and family formation, assumed for simplicity to coincide at age 25.
\citet{fischer73} pioneered numeric studies of life cycle models of the demand for life insurance and life annuities when loads are of the one-parameter variety.

\begin{figure}[htb]
\begin{center}
\includegraphics{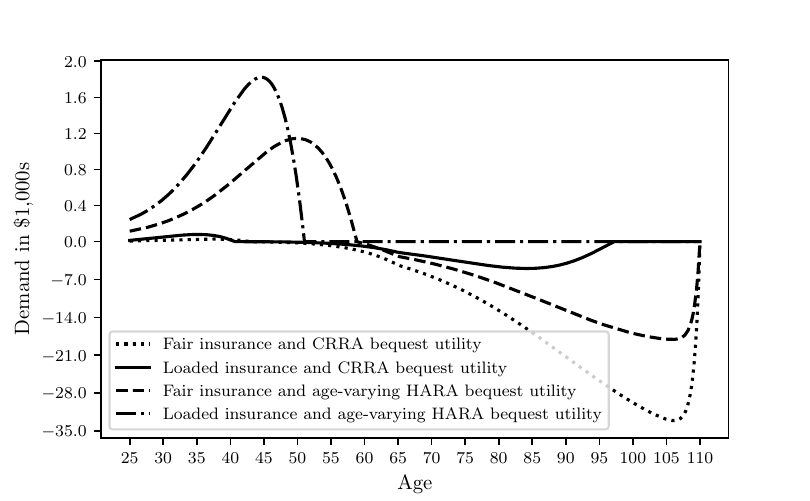}
\vspace*{-1.0cm}
\caption{Optimal life insurance and life annuity purchase over the life cycle.}
\label{fig:fullcycle}
\end{center}
\end{figure}

Bequest utility here is either CRRA (normal bequests) or HARA with an age-varying shift parameter (necessity bequests that transition to luxury bequests).
Figure \ref{fig:fullcycle} portrays the case of an 18$\%$ load on both life insurance and life annuities, and compares it to the case of zero loads.
Positive annuity loads and age-varying bequest jointly modify the demand for life-contingent insurance over the life cycle, to the point where decades-long spans of non-participation open up.
In more detail, fair insurance ensures continuous participation regardless of whether bequest utility is CRRA or age-varying HARA.

A load of 18$\%$ combined with CRRA utility sees two periods of non-participation, one running from age 41 to 44, and the other beginning at age 97.
Intuition for the midlife non-participation period can be gained from observing that fair insurance combined with CRRA utility sees the crossover from life insurance to life annuities occur between ages 41 and 44.
Once the load is introduced, non-participation seeps out from either side of the fair-insurance cutpoint.
Similarly, for the late-life non-participation period, non-participation spreads to the left from the maximum-lifespan cut-point, namely, age 110.

Loaded insurance and age-varying bequests together see lifelong non-participation from age 51.  
It is evident that realistic loads can have strong effects, especially on annuity demands in conjunction with luxury bequests for retirees.
Moreover, the demand for life insurance is weak when insurance is not a necessity during some working years.
The figure also shows that age-varying bequest motives combine with loads to explain the deep annuity puzzle.

\begin{figure}[htb]
\begin{center}
\includegraphics{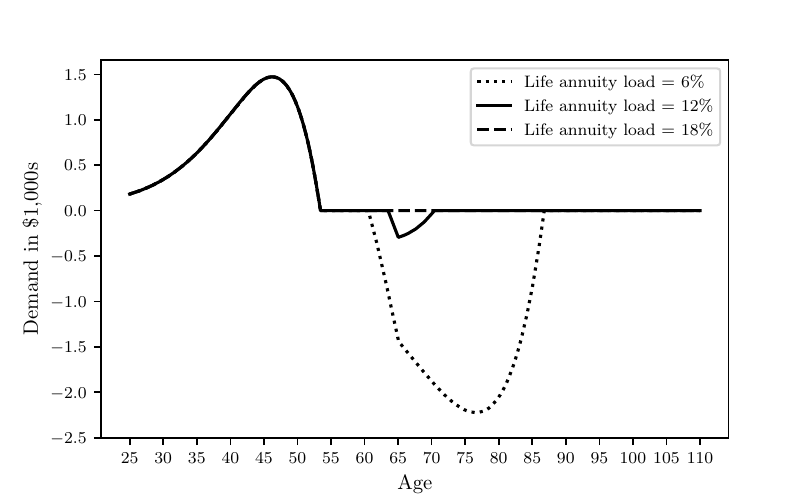}
\vspace*{-1.00cm}
\caption{Effects of different annuity loads on the demand for life insurance}
\label{fig:feedforward}
\end{center}
\end{figure}

Next, we examine the effects of annuity loads on the demand for life insurance.
In particular, we consider loads of either 6$\%$, 12$\%$ or 18$\%$ while holding the life insurance load fixed at 12$\%$.
Bequest utility is age-varying HARA.
Figure \ref{fig:feedforward} refers.
Our computations suggest that there is no effect of changes in annuity loads on the demand for life insurance.
For example, when the annuity load is raised from 12$\%$ to 18$\%$, there is a fall in the demand for life annuities accompanied by a lengthening of the midlife non-participation period.
However, the age at which non-participation begins remains unchanged at age 53.
Likewise, there is no change in the life-insurance profile.
Part of the intuition is that life-contingent insurance here is of the instantaneous-term variety, so that individuals only look an instant ahead when making insurance decisions.

\begin{figure}[htb]
\begin{center}
\includegraphics{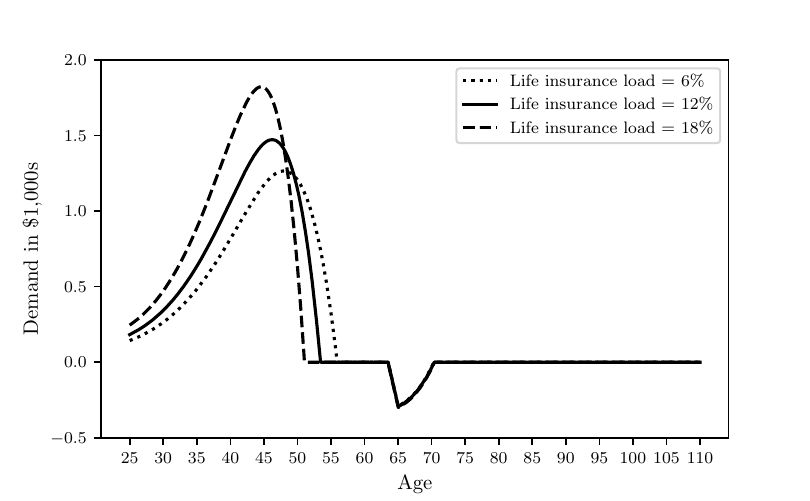}
\vspace*{-1.00cm}
\caption{Effects of different insurance loads on the demand for life annuities}
\label{fig:feedforward2}
\end{center}
\end{figure}

Finally, consider the effects of changes in life insurance loads on the demand for life annuities.
Bequest utility is again age-varying HARA.
Figure \ref{fig:feedforward2} refers.
Paralleling the results portrayed in Figure \ref{fig:feedforward}, there is no feedback effect, regardless of life insurance loads.
Take the case of a 12$\%$ load.
The corresponding annuitization phase begins at age 64 and ends at age 70.
It is not affected by changes in life insurance loads.
This figure also shows that the demand for life insurance is not highly sensitive to loads, in contrast to the demand for annuities.

\section{Conclusions}
\label{sec:conclusions}
Numerous previous investigations have examined the optimal demand for life insurance and life annuities in life cycle settings and under insurance loads.
Loads have been modelled by means of a single parameter that effectively eliminates bid-ask spreads in insurance markets.
This helps explain why numeric investigations have typically predicted continuous participation in insurance markets along with strong demand for annuities.
A two-parameter model can restore positive bid-ask spreads.
Positive spreads render markets incomplete, making it necessary to examine optimal insurance purchases by means of computations.

Our computations suggest that positive annuity loads induce up to two periods of non-participation, one in midlife and the other adjoining the maximum age at death.
Luxury bequests are not a necessary condition for two (rather than one) non-participation periods.

When bequests are luxuries, loads can have big effects on annuity demands.
According to our computations for someone retiring at age 65, if the load is 14$\%$ or more of premiums then the demand for annuities is negligible.
More generally, our computations suggest that the demand for loaded annuities is comparatively weak around age 65 even when bequests are not luxuries.
By contrast, the demand for fair annuities is comparatively strong even when bequests are luxuries, especially in advanced old age.

To the extent bequests are necessities, loads no longer have large negative effects on insurance demands.
Life insurance is a case in point.
According to our computations for a 25 year old, for whom bequests are a necessity (while in gradual transit to being luxuries), the demand for life insurance (calibrated by reference to an 18$\%$ load at age 65) is strong.
Moreover, compared to fair insurance, a life insurance load of 18$\%$ (calibrated to age 65) sees a higher peak demand for life insurance.
This peak occurs at age 45.
In other words, life insurance is a Giffen good in mid working life.
This echoes a numeric finding due to \citet{pliska07} -- see their Figures 5 and 6.
According to our Figure \ref{fig:feedforward2}, an 18$\%$ load sees the demand for life insurance peter out at age 51.
Moreover, the same load on annuities sees a negligible demand for them.
A maximum lifespan of 110 years therefore implies that the period of non-participation in the market for life-contingent insurance could be up to 59 years in length.

Our life-cycle computations suggest that changes in annuity loads do not affect life insurance decisions, although such changes do alter the age at which midlife non-participation ends.
Likewise, changes in life insurance loads do not affect the demand for annuities.  

Our computations need to be interpreted with caution.
First, post-retirement parameters were based on \citet{lockwood18}, who derives his estimates from the Health and Retirement Study.
As a consequence of Social Security, there is some pre-annuitization of retirement wealth.
Second, adverse-selection issues make instantaneous-term products difficult for annuity providers, especially in the case of potential customers at advanced ages.
Finally, there is ample scope for further investigation of bequest motives during working life.
These will vary from one household to the next.
For example, our assumed age-dependency profile attained a minimum at age 40, corresponding to a maximum desire to provide for dependents.
Yet this aspect of preferences is highly idiosyncratic across households.
Estimating the bequest-utility parameters of working-age people is a topic we leave to future research.

\section*{Acknowledgements}
Pavel Shevchenko acknowledges the support of Australian Research Council's Discovery Projects funding scheme (project number DP160103489).
Aleksandar Arandjelovi{\'c} acknowledges support from the  International Cotutelle Macquarie University Research Excellence Scholarship.
Lance Fisher made constructive comments, as did participants of the 26th International Congress on Insurance: Mathematics and Economics, and two anonymous referees.

\appendix
\section{Truncated life cycle}
\label{ap:PY}
Some investigators focusing on life insurance collapse the retirement phase into a single point in time.
Numeric studies with single-parameter models of loads and financial plans terminating with the individual’s retirement (rather than her maximum lifespan) were pioneered by \citet{pliska07}.
They found that (subsidized) annuity demands over this truncated horizon were modest, though discernible.
Participation was found to be continuous.
We redo their Figure 6, including here the case of positive (rather than negative) annuity loads.

\begin{figure}[htb]
\begin{center}
\includegraphics{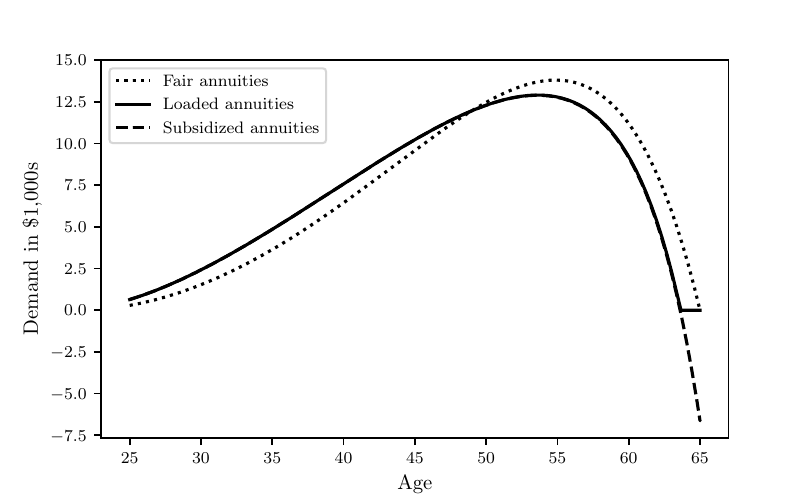}
\vspace*{-1.0cm}
\caption{Positive annuity loads in the Pliska--Ye model}
\label{fig:PY}
\end{center}
\end{figure}

In Figure \ref{fig:PY}, the planning horizon can be seen from the horizontal axis to be 40 years.
Demands for life insurance and life annuities are shown on the vertical axis.
The dotted line portrays the base case of Pliska and Ye's numeric analysis: loads are zero, the coefficient of relative risk aversion is 4, the rate of time preference is 3$\%$ per year, and the discount rate is 4$\%$ per year.
The hazard rate is assumed to be a simple linear function of age -- see \citet{pliska07} for details.
There is no annuity phase in the base case.
The dashed line shows demands under a single-parameter model of the premium-insurance ratio.
That case corresponds to $\kappa_{\mathrm{ins}}=4$ and $\kappa_{\mathrm{ann}}=1/4$.
Introducing subsidized annuities (along with loaded life insurance) induces a positive annuity phase.
It is about one year in length.
Finally, the solid line shows demands under a two-parameter model, thereby ensuring that life insurance and life annuities both carry loads.
The premium-insurance ratio corresponds to $\kappa_{\mathrm{ins}}=4$ during the insurance phase and $\kappa_{\mathrm{ann}}=4$ thereafter.
In place of an annuity phase there is a non-participation period.
It is about one year in length and adjoins the retirement time $T=40$.
As with zero loads, there is no annuity phase in this case.

\section{Estimating a Gompertz mortality model}
\label{ap:mortrate}
We calibrate a model for the mortality rate $\lambda(t)$ to data.
Our sample consists of life tables from the 2019 cohort of the G12 countries (Australia, Belgium, Canada, France, Germany, Italy, Japan, Netherlands, Spain, Sweden, Switzerland, UK and USA), which was obtained from the \citet{hmd22}.
For each country $i=1, 2, \hdots, 13$, and age $x=25, 26, \hdots, 110$, we have data on $l_{x}^{i}$, the number of survivors at age $x$, and $d_{x}^{i}$, the number of deaths within the subsequent year from those survivors aged $x$ years.
The age 110 is not a single-year age, being treated as a cemetery state with $l_{110}^{i} = d_{110}^{i}$.
We aggregate both the number of survivors and the number of deaths over all countries, and denote the corresponding age-dependent aggregates $l_{x}$ and $d_{x}$, respectively.

We aim to estimate the parameters $b$ and $m$ of the mortality model set out in Table \ref{tab:base_params}.
We proceed as follows:
\begin{itemize}
\item
Approximate the survival function via the Levenberg--Marquardt nonlinear least-squares method;
\item
Calibrate the mortality rate via the maximum likelihood method.
\end{itemize}
Both the survival function as well as the mortality rate enter directly into the definition of the value function $V$ according to the dynamic programming principle in \ref{ap:DP} below, so that it is important to have a good estimate for both functions.

\subsection{Fitting the survival function}
Under the parametric model for $\lambda(t)$ set out in Table \ref{tab:base_params}, the survival function $\bar{F}(t)$ takes the form
\begin{equation*}
\bar{F}(t) = %
\exp\Bigl(-\exp\bigl(\tfrac{25-m}{b}\bigr)\bigl( \exp\bigl(\tfrac{t}{b}\bigr) - 1\bigr)\Bigr).
\end{equation*}
It admits the representation $\bar{F}(t) = l_{25+t} / l_{25}$.
We have information about the aggregated number of survivors $l_{25+t}$ for $t=0, 1, \hdots, 85$.
We thus have $86$ training points and aim to find parameters $\hat{b}$ and $\hat{m}$ such that the error
\begin{equation*}
\frac{1}{2} \sum_{t=0}^{85} \Bigl( \bar{F}(t) - \frac{l_{25+t}}{l_{25}} \Bigr)^{2}
\end{equation*}
is minimized.
This is done via the Levenberg--Marquardt nonlinear least squares method; see Table \ref{tab:GMparam} for the estimated coefficients.

\subsection{Maximum likelihood estimation}
We follow \citet{lenart14} and assume $d_{25+t}$ follows a Poisson distribution,
\begin{equation*}
d_{25+t} \sim \mathrm{Poisson}(E_{25+t}\,\lambda(25+t)),
\end{equation*}
where $E_{25+t}$ denotes the total number of person-years exposed to death at age $25+t$, defined analytically via $E_{25+t} = \int_{0}^{1} l_{25+t+u}\, \mathrm{d}u$.
By construction it follows that $l_{25+t} \ge E_{25+t} \ge l_{25+t+1}$ for all $t$.
$E_{25+t}$ accounts for the fact that not all $l_{25+t}$ lives are exposed to mortality risk throughout the whole period $[25+t, 25+t+1)$, but only for a sub-period until death.

For simplicity, we assume that
\begin{equation*}
E_{25+t} = l_{25+t} - (1-a_{25+t})d_{25+t}, \quad t=0, 1, ..., 84,
\end{equation*}
where $a_{25+t}$ denotes the average number of years lived within the age-interval $[25+t, 25+t+1)$ for those people dying at that age, which we set to $a_{25+t} = 0.5$ for $t=0, 1, ..., 84$.
This is in line with the methodology used for the construction of life tables in the \citet{hmd22}.

As already mentioned, the age $x=110$ is special in our sample, because it is not a single-year age, but a cemetery state, where $l_{110} = d_{110}$.
In order to avoid introducing a special treatment and therefore a modelling bias, we exclude this state for the remainder of this subsection, and estimate our parameters only based on the ages $x=25, 26, ..., 109$.
Note that using the cemetery age $x=110$ for approximating the survival function in the previous subsection is not a problem though, because $l_{110}$ still gives the proper interpretation of those lived at the beginning of the cemetery age.

We re-parameterize the mortality rate,
\begin{equation*}
\lambda(t) = \tfrac{1}{b} \exp\bigl( \tfrac{25-m}{b} \bigr) \exp\bigl( \tfrac{t}{b} \bigr) = \alpha \exp(\beta t), 
\end{equation*}
where $\alpha = \exp((25-m)/b)/b$ and $\beta = 1/b$.
Following \citet{lenart14}, the maximum likelihood estimator $\hat{\alpha}$ for $\alpha$ is
\begin{equation*}
\hat{\alpha} = \frac{\sum_{t=0}^{84}d_{25+t}}{\sum_{t=0}^{84}E_{25+t}\exp(\beta t)},
\end{equation*}
and the maximum likelihood estimator $\hat{\beta}$ for $\beta$ is the root of the function
\begin{equation*}
f(y) = \frac{\sum_{t=0}^{84}d_{25+t}t}{\sum_{t=0}^{84}d_{25+t}} - \frac{\sum_{t=0}^{84}E_{25+t}\exp{(yt)}t}{\sum_{t=0}^{84}E_{25+t}\exp{(yt)}}.
\end{equation*}
Conveniently, this representation allows for a straightforward calculation of the first derivative,
\begin{equation*}
f'(y) = \Biggl(\frac{\sum_{t=0}^{84}E_{25+t}\exp{(yt)}t}{\sum_{t=0}^{84}E_{25+t}\exp{(yt)}}\Biggr)^{2} - \frac{\sum_{t=0}^{84}E_{25+t}\exp{(yt)}t^{2}}{\sum_{t=0}^{84}E_{25+t}\exp{(yt)}},
\end{equation*}
and we can find the root of $f$ numerically, as suggested by \citet{lenart14}, via the Newton--Raphson method; see Table \ref{tab:GMparam} for the estimated coefficients.

\subsection{Numerical calibration}
We estimate the Gompertz mortality parameters $b$ and $m$ via both the Levenberg--Marquardt and the maximum likelihood method.
For the final model coefficients, we choose a weighted average, where $25\%$ of the weight is allocated to the first method (fitting the survival function), and $75\%$ is allocated to the second method (maximum likelihood estimation).
The reason for this weighting is that for our optimal control problem, both the survival function and the mortality rate appear in the definition of the value function.
However, since the mortality rate $\lambda(t)$ plays a more prominent role through the definition of the premium-insurance ratios $\eta(t)$ and $\theta(t)$, we overweight the second set of estimated parameters.

\begin{table}[htb]
\caption{Estimated Gompertz model parameters}
\begin{center}
\begin{tabular}{@{}ccc@{}}
\toprule
Method & $\hat{b}$ & $\hat{m}$\\
\midrule
Levenberg--Marquardt & $9.45$ & $88.79$ \\
Maximum likelihood & $9.35$ & $88.05$ \\
\midrule
Weighted average & $9.38$ & $88.23$\\
\bottomrule
\end{tabular}
\label{tab:GMparam}
\end{center}
{\small Note: Reported values are rounded to two decimal places.\par}
\end{table}

\begin{figure}[htb]
\begin{center}
\includegraphics{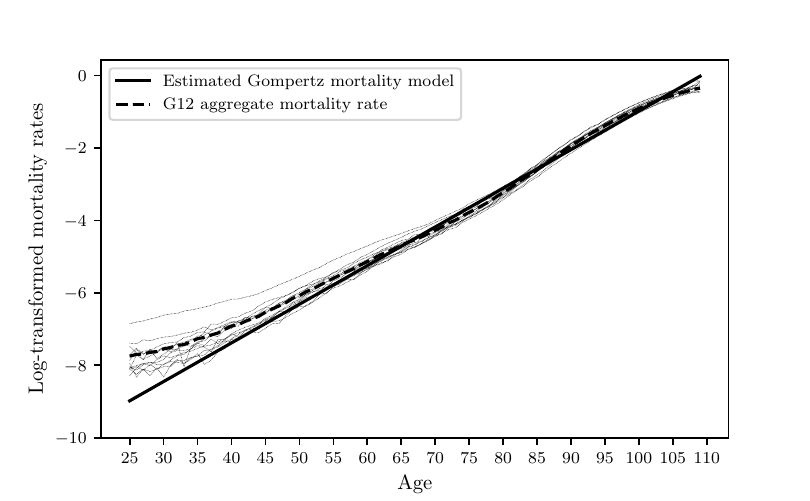}
\vspace*{-1.0cm}
\caption{Mortality rates of G12 countries, their aggregate and Gompertz model fit}
\label{fig:mort_rate}
\end{center}
\end{figure}

Figure \ref{fig:mort_rate} portrays our estimated model, the aggregated mortality rate, and the mortality rates of each individual constituent country of the G12 group.
The mortality rates are defined via $m_{x}^{i} = d_{x}^{i} / E_{x}^{i}$ and $m_{x} = d_{x} / E_{x}$, where $E_{x}^{i}$, the country-level total number of person-years exposed to death at age $x$ were obtained similarly as above by replacing the aggregate values by country-level values.

\section{Age-earnings and age-dependency profiles}
\label{ap:aead}
We follow \citet{mincer74}, who discusses a quadratic form of the income function fit to the logarithm of weekly earnings.
All dollar values in this section were rounded generously.

The average income of single retirees in Lockwood's (2018) sample is $\$$18{,}360, expressed in constant 2010 dollars using the Consumer Price Index for Urban Wage Earners and Clerical Workers (CPI-W).
The annual average CPI-W stood at 213.967 points in 2010 and 283.926 points in 2022 (average over the first half of the year).
This corresponds to an increase of approximately 32.69$\%$.
Adjusting Lockwood's income figure, we therefore assume a post-retirement income of $\$$24{,}360.
At the moment of retirement, we postulate a drop in income of 60$\%$, which implies that income immediately pre-retirement is $\$$60{,}900.

To generate more data points, we assume that income grows by $35$\% in the first 10 years in the labor market, by $25$\% in the next 15 years, and declines by $15$\% in the final 15 years \citep{murphy90}.
We thus arrive at the following data points:
\begin{itemize}
\item
The initial income is $\$$42{,}460;
\item
After 10 years in the labor market, the income is $\$$57{,}320;
\item
The peak occurs after 25 years with an income of $\$$71{,}650;
\item
Immediately before retirement, the income is $\$$60{,}900.
\end{itemize}

We fit an age-income profile $y(t)$ to these data points, leading to
\begin{equation*}
y(t) = \exp\bigl(-0.000763\, t^{2} + 0.0398\, t + 10.65 \bigr).
\end{equation*}

Turning to the age-dependency profile, Lockwood estimates $c_{b}(t)$ post-retirement to be $\$$24{,}800.
Adjusting Lockwood's value to constant 2022 dollars yields the value of $\$$32{,}900.
Our post-retirement income is $\$$24{,}360.
Therefore, $c_{b}(t)$ post-retirement is approximately $35$\% larger than post-retirement income.
For symmetry reasons, we postulate that pre-retirement, minimum bequest after 20 years in the labor market should be approximately $35$\% larger than the income at that time.
Note that, due to our choice of utility function $B(t,z)$ for the case $c_{b}(t) < 0$, the minimum bequest is $- \phi/(1-\phi) c_{b}(t)$, so we have to account for the scaling with $\bar{\phi} = \phi/(1-\phi)$.
We thus arrive at the following data points: $c_{b}(0) = 0$, $c_{b}(20) = -1.35\, y(20) / \bar{\phi}$, and $c_{b}(40) = 32{,}900$.

We fit piecewise cubic splines over the intervals $[0, 20]$ and $[20, 40]$ with natural boundary conditions, meaning that the first derivatives are set to zero at the boundaries of those two intervals.
This yields approximately the following specification for $c_{b}(t)$, where for ease of presentation we rounded the coefficients to two decimal places:
\begin{equation*}
c_{b}(t) = 
\begin{cases}
4.31\, t^{3} - 98.42\, t^{2} & t \in [0, 20),\\
\begin{aligned}
&-6.36\, (t-20)^{3} +160.11\, (t-20)^{2}\\
&\hspace{2cm} + 1{,}233.75\, (t-20) - 4{,}897.43
\end{aligned} & t \in [20, 40),\\
32{,}900 & t \in [40, 85].
\end{cases}
\end{equation*}

\section{Dynamic programming}
\label{ap:DP}
For completeness, we set out the dynamic program, leading to a numerical algorithm that can be used to compute optimal solutions; see Algorithm \ref{dp_algo}.
Here we treat a problem of deterministic optimal control.
For a textbook treatment, see \citet{fleming06}.

In assuming a Gompertz mortality model for $\lambda(t)$ as specified in Table \ref{tab:base_params}, the density $f$ of the individual's remaining lifetime $\tau$ has the form
\begin{equation*}
f(t) = \lambda(t) \bar{F}(t) = \frac{1}{b}\exp\Bigl(C + \ln(C) + \tfrac{t}{b} - C\exp\bigl(\tfrac{t}{b}\bigr)\Bigr),
\end{equation*}
where $C = \exp((25-m)/b)$.
From this we can see that $f(t) \to 0$ as $t \to \infty$, because the term $C\exp(t/b)$ dominates the term $t/b$ inside the exponent.
However, for each $t > 0$, we also have $f(t) > 0$ as well as $\bar{F}(t) > 0$.
In other words, assuming a Gompertz mortality model for $\lambda(t)$ implies that for any time $t > 0$, there is a positive probability that the individual will survive up to this time.

\citet{richard75} avoided this case by assuming that the support of the distribution of $\tau$ is bounded from above by a finite constant, say $\tilde{T} > 0$.
Choosing $T = \tilde{T}$ would then yield a proper life cycle model.
One disadvantage of this approach is that one cannot use the Gompertz law of mortality, which is popular in actuarial practice.
Alternatively, one could truncate the distribution of $\tau$ to a finite interval $[0, \tilde{T}]$ by replacing the density $f(t)$ above with $\tilde{f}(t) = f(t) / F(\tilde{T}) \mathrm{1}_{[0, \tilde{T}]}(t)$, where $F(t)$ denotes the cumulative distribution function of $\tau$.
In that case, we would have
\begin{equation*}
\mathbb{P}(\{ \tau \le \tilde{T} \}) = \int_{0}^{\tilde{T}} \tilde{f}(t)\, \mathrm{d}t = \frac{F(\tilde{T})}{F(\tilde{T})} = 1.
\end{equation*}

It would be possible to replace the Gompertz mortality model by another model satisfying $\bar{F}(t) = 0$ for all $t$ above a reasonable time $\tilde{T}$, say $\tilde{T} = 85$, or by truncating the model as described above to obtain a distribution over $[0, \tilde{T}]$.
However, we note that for the chosen model parameters as specified in Table \ref{tab:base_params}, the probability that a 25 year old individual survives past the year 110 is $\bar{F}(85) \approx 3.78 \cdot 10^{-5}$.
Therefore, the Gompertz mortality model with parameters as described in Table \ref{tab:base_params} and with $T = 85$ is an adequate approximation to optimal consumption and premium plans over the full life cycle.
This approximation omits only survival scenarios with a negligible probability of occurrence.

Given an initial level of wealth $w$, time $s < T$ as well as a pair $(c,p)$ of consumption and insurance plans, let
\begin{multline*}
J(s,w;c,p) = \mathbb{E}\Bigl[ \int_{s}^{\tau \wedge T} \mathrm{e}^{-\beta (t-s)}U(c(t))\, \mathrm{d}t + \mathrm{e}^{-\beta (\tau-s)}B(\tau, Z(\tau))\mathrm{1}_{\{\tau \le T\}} \\ + \mathrm{e}^{-\beta (T-s)}B(T, W(T))\mathrm{1}_{\{\tau > T\}} \, {\big|}\, \tau > s\Bigr],
\end{multline*}
where $\mathbb{E}[\,\cdot\, |\, \tau > s]$ denotes the expectation conditional on $\{ \tau > s \}$.
The only source of randomness is the uncertain lifetime $\tau$.
Following an approach pioneered by \citet{yaari65}, $J(s,w;c,p)$ can be rewritten:
\begin{multline*}
J(s,w;c,p) = \int_{s}^{T}\mathrm{e}^{-\beta (t-s)}\bigl( \bar{F}(t,s)U(c(t)) + f(t,s)B(t,Z(t)) \bigr)\, \mathrm{d}t \\ + \mathrm{e}^{-\beta (T-s)}\bar{F}(T,s)B(T,W(T)),
\end{multline*}
where $\bar{F}(t,s) = \bar{F}(t) / \bar{F}(s)$ denotes the conditional survival probability, and $f(t,s) = f(t) / \bar{F}(s) $ denotes the conditional probability density.

We can now use deterministic dynamic programming over the fixed time horizon $[0, T]$.
We assume that $y(t)$ is integrable (which is the case for the age-income profile specified in \ref{ap:aead}).
Moreover, we restrict our attention to consumption and premium plans $(c,p)$ which are bounded and measurable.
Then Equation \ref{eq:wealth} is well defined.
To further assure admissibility, for each $w$ and $s \in [0, T)$, let $\mathcal{A}(s,w)$ denote the set of all pairs $(c,p)$ of consumption and premium plans on $[s, T]$, with
\begin{enumerate}
\item
$c(t) > 0$ for all $t \in [s, T]$;
\item
$Z(t) > \max\{0, - \tfrac{\phi}{1-\phi}c_{b}(t)\}$ for all $t \in [s, T)$;
\item
$W(T) > \max\{0, - \tfrac{\phi}{1-\phi}c_{b}(T)\}$.
\end{enumerate}

The value function is then defined via $V(s,w) = \sup_{(c,p) \in \mathcal{A}(s,w)} J(s,w;c,p)$ and the optimization problem \ref{eq:optim} can equivalently be written as $V(w) = V(0,w)$.
Note that we do not need to define the set $\mathcal{A}(T,w)$ because the terminal boundary condition is known: $V(T,w) = B(T, w)$.

We now formulate the dynamic programming principle (DPP).
For all $s < t$ in $[0, T]$ and $w$,
\begin{multline*}
V(s,w) = \sup_{(c,p) \in \mathcal{A}(s,w)} \Bigl( \mathrm{e}^{-\beta (t-s)}\bar{F}(t,s) V(t, W(t)) \\ + \int_{s}^{t}\mathrm{e}^{-\beta (u-s)}\bigl( \bar{F}(u,s)U(c(u)) + f(u,s)B(u,Z(u)) \bigr)\, \mathrm{d}u \Bigr).\end{multline*}
The DPP gives rise to a numerical algorithm which we use to solve the optimization problem, see Algorithm \ref{dp_algo} below.

\clearpage
\begin{algorithm}[h!]
{\footnotesize{
\caption{Dynamic programming}\label{dp_algo}
\begin{algorithmic}[1]
\State
Select a time grid $0=t_{0}<t_{1}<\hdots<t_{N-1}<t_{N}=T$ with steps $\Delta t$;
\State
Select (possibly time-dependent) nodes for the state variable $W(t)$: $w_{j}$, $j=1,\hdots,J$;
\State
Initialise $\hat{V}(t_{N}, w_{j})=B(t_{N}, w_{j})$ for $j=1,\hdots,J$;
\For{$i = N-1, \hdots, 0$}
\State
Interpolate/extrapolate $\hat{V}(t_{i+1},w_{j})$, $j=1,\hdots,J$ to approximate $\hat{V}(t_{i+1},w)$
\State
for any $w$ (for $i=N-1$ the maturity condition is known: $\hat{V}(T,w) = B(T,w)$);
\State
Calculate $a_{i}^{(1)} = \int_{t_{i}}^{t_{i+1}}\mathrm{e}^{-\beta(s-t_{i})}\bar{F}(s,t_{i})\, \mathrm{d}s$, $a_{i}^{(2)} = \int_{t_{i}}^{t_{i+1}}\mathrm{e}^{-\beta(s-t_{i})}f(s,t_{i})\, \mathrm{d}s$
\State
and $a_{i}^{(3)} = \mathrm{e}^{-\beta(t_{i+1}-t_{i})}\bar{F}(t_{i+1}, t_{i})$, e.g. via numerical integration procedure;
\For{$j=1,\hdots,J$}
\State
$\hat{V}(t_{i},w_{j})=\sup_{(c,p)}\Bigl(a_{i}^{(1)}U(c) + a_{i}^{(2)}B(t_{i}, Z(t_{i})) + a_{i}^{(3)}\hat{V}(t_{i+1}, \tilde{w})\Bigr)$,
\State
where $\tilde{w}=w_{j}(1+r\Delta t)+(y(t_{i})-c-p)\Delta t$;
\EndFor
\EndFor
\State Calculate forward in time optimal trajectory
\State Set initial wealth, e.g. $W(0)=0$
\For{${i} = 0, \hdots, N-1$}
\State Find optimal controls
\State $(c^\ast(t_{i}),p^\ast(t_{i})) = \arg\sup_{(c,p)}\Bigl(a_{i}^{(1)}U(c)+a_{i}^{(2)}B(t_{i}, Z(t_{i}))+
a_{i}^{(3)}\hat{V}(t_{i+1}, \tilde{w}) \Bigr)$,
\State where $\tilde{w}=W(t_{i})(1+r\Delta t)+(y(t_{i})-c-p)\Delta t$;
\State Find wealth at the next time: $W(t_{i+1})=W(t_{i})(1+r\Delta t)+(y(t_{i})-c^\ast(t_{i})-p^\ast(t_{i}))\Delta t$.
\EndFor
\State Calculated $(c^\ast(t_{i}),p^\ast(t_{i}))$ and $W(t_i)$ are optimal trajectories for consumption, insurance and wealth.
\end{algorithmic}
}}
\end{algorithm}
\clearpage

\bibliographystyle{elsarticle-harv} 
\bibliography{main}

\begin{thebibliography}{25}
\expandafter\ifx\csname natexlab\endcsname\relax\def\natexlab#1{#1}\fi
\providecommand{\url}[1]{\texttt{#1}}
\providecommand{\href}[2]{#2}
\providecommand{\path}[1]{#1}
\providecommand{\DOIprefix}{doi:}
\providecommand{\ArXivprefix}{arXiv:}
\providecommand{\URLprefix}{URL: }
\providecommand{\Pubmedprefix}{pmid:}
\providecommand{\doi}[1]{\href{http://dx.doi.org/#1}{\path{#1}}}
\providecommand{\Pubmed}[1]{\href{pmid:#1}{\path{#1}}}
\providecommand{\bibinfo}[2]{#2}
\ifx\xfnm\relax \def\xfnm[#1]{\unskip,\space#1}\fi
\bibitem[{Brown and Finkelstein(2007)}]{brown07}
\bibinfo{author}{Brown, J.}, \bibinfo{author}{Finkelstein, A.},
  \bibinfo{year}{2007}.
\newblock \bibinfo{title}{Why is the market for long-term care insurance so
  small?}
\newblock \bibinfo{journal}{Journal of Public Economics} \bibinfo{volume}{91},
  \bibinfo{pages}{1967--1991}.
\bibitem[{Carroll(1998)}]{carroll98}
\bibinfo{author}{Carroll, C.}, \bibinfo{year}{1998}.
\newblock \bibinfo{title}{Why Do the Rich Save So Much?}
\newblock \bibinfo{type}{Working Paper} \bibinfo{number}{6549}. National Bureau
  of Economic Research (NBER).
\bibitem[{Davis and Norman(1990)}]{davis90}
\bibinfo{author}{Davis, M.}, \bibinfo{author}{Norman, A.},
  \bibinfo{year}{1990}.
\newblock \bibinfo{title}{Portfolio selection with transaction costs}.
\newblock \bibinfo{journal}{Mathematics of Operations Research}
  \bibinfo{volume}{15}, \bibinfo{pages}{676--713}.
\bibitem[{De~Nardi et~al.(2010)De~Nardi, French and Jones}]{denardi10}
\bibinfo{author}{De~Nardi, M.}, \bibinfo{author}{French, E.},
  \bibinfo{author}{Jones, J.}, \bibinfo{year}{2010}.
\newblock \bibinfo{title}{Why do the elderly save? {T}he role of medical
  expenses}.
\newblock \bibinfo{journal}{Journal of Political Economy}
  \bibinfo{volume}{118}, \bibinfo{pages}{39--75}.
\bibitem[{Feigenbaum et~al.(2013)Feigenbaum, Gahramanov and
  Tang}]{feigenbaum2013really}
\bibinfo{author}{Feigenbaum, J.}, \bibinfo{author}{Gahramanov, E.},
  \bibinfo{author}{Tang, X.}, \bibinfo{year}{2013}.
\newblock \bibinfo{title}{Is it really good to annuitize?}
\newblock \bibinfo{journal}{Journal of Economic Behavior \& Organization}
  \bibinfo{volume}{93}, \bibinfo{pages}{116--140}.
\bibitem[{Fischer(1973)}]{fischer73}
\bibinfo{author}{Fischer, S.}, \bibinfo{year}{1973}.
\newblock \bibinfo{title}{A life cycle model of life insurance purchases}.
\newblock \bibinfo{journal}{International Economic Review}
  \bibinfo{volume}{14}, \bibinfo{pages}{132--152}.
\bibitem[{Fleming and Soner(2006)}]{fleming06}
\bibinfo{author}{Fleming, W.}, \bibinfo{author}{Soner, H.},
  \bibinfo{year}{2006}.
\newblock \bibinfo{title}{Controlled {M}arkov Processes and Viscosity
  Solutions}. volume~\bibinfo{volume}{25} of
  \textit{\bibinfo{series}{Stochastic Modelling and Applied Probability}}.
\newblock \bibinfo{edition}{Second} ed., \bibinfo{publisher}{Springer, New
  York}.
\bibitem[{Hakansson(1969)}]{hakansson69}
\bibinfo{author}{Hakansson, N.}, \bibinfo{year}{1969}.
\newblock \bibinfo{title}{Optimal investment and consumption strategies under
  risk, an uncertain lifetime, and insurance}.
\newblock \bibinfo{journal}{International Economic Review}
  \bibinfo{volume}{10}, \bibinfo{pages}{443--466}.
\bibitem[{Heijdra et~al.(2014)Heijdra, Mierau and
  Reijnders}]{heijdra2014tragedy}
\bibinfo{author}{Heijdra, B.J.}, \bibinfo{author}{Mierau, J.O.},
  \bibinfo{author}{Reijnders, S.}, \bibinfo{year}{2014}.
\newblock \bibinfo{title}{A tragedy of annuitization? {L}ongevity insurance in
  general equilibrium}.
\newblock \bibinfo{journal}{Macroeconomic Dynamics} \bibinfo{volume}{18},
  \bibinfo{pages}{1607--1634}.
\bibitem[{Huang and Milevsky(2008)}]{huang08b}
\bibinfo{author}{Huang, H.}, \bibinfo{author}{Milevsky, M.},
  \bibinfo{year}{2008}.
\newblock \bibinfo{title}{Portfolio choice and mortality-contingent claims:
  {T}he general {HARA} case}.
\newblock \bibinfo{journal}{Journal of Banking and Finance}
  \bibinfo{volume}{32}, \bibinfo{pages}{2444--2452}.
\bibitem[{{Human Mortality Database}(2022)}]{hmd22}
\bibinfo{author}{{Human Mortality Database}}, \bibinfo{year}{2022}.
\newblock \URLprefix \url{www.mortality.org}. \bibinfo{note}{{D}ata downloaded
  on 13 December 2022. Institutional sponsors: Max Planck Institute for
  Demographic Research (Germany), University of California, Berkeley (USA), and
  French Institute for Demographic Studies (France)}.
\bibitem[{Lenart(2014)}]{lenart14}
\bibinfo{author}{Lenart, A.}, \bibinfo{year}{2014}.
\newblock \bibinfo{title}{The moments of the {G}ompertz distribution and
  maximum likelihood estimation of its parameters}.
\newblock \bibinfo{journal}{Scandinavian Actuarial Journal}
  \bibinfo{volume}{3}, \bibinfo{pages}{255--277}.
\bibitem[{Lockwood(2012)}]{lockwood12}
\bibinfo{author}{Lockwood, L.}, \bibinfo{year}{2012}.
\newblock \bibinfo{title}{Bequest motives and the annuity puzzle}.
\newblock \bibinfo{journal}{Review of Economic Dynamics} \bibinfo{volume}{15},
  \bibinfo{pages}{226--243}.
\bibitem[{Lockwood(2018)}]{lockwood18}
\bibinfo{author}{Lockwood, L.}, \bibinfo{year}{2018}.
\newblock \bibinfo{title}{Incidental bequests and the choice to self-insure
  late-life risks}.
\newblock \bibinfo{journal}{American Economic Review} \bibinfo{volume}{108},
  \bibinfo{pages}{2513--2550}.
\bibitem[{Magill and Constantinides(1976)}]{magill76}
\bibinfo{author}{Magill, M.}, \bibinfo{author}{Constantinides, G.},
  \bibinfo{year}{1976}.
\newblock \bibinfo{title}{Portfolio selection with transactions costs}.
\newblock \bibinfo{journal}{Journal of Economic Theory} \bibinfo{volume}{13},
  \bibinfo{pages}{245--263}.
\bibitem[{Merton(1971)}]{merton71}
\bibinfo{author}{Merton, R.}, \bibinfo{year}{1971}.
\newblock \bibinfo{title}{Optimum consumption and portfolio rules in a
  continuous-time model}.
\newblock \bibinfo{journal}{Journal of Economic Theory} \bibinfo{volume}{3},
  \bibinfo{pages}{373--413}.
\bibitem[{Milgram(1985)}]{milgram85}
\bibinfo{author}{Milgram, M.}, \bibinfo{year}{1985}.
\newblock \bibinfo{title}{The generalized integro-exponential function}.
\newblock \bibinfo{journal}{Mathematics of Computation} \bibinfo{volume}{44},
  \bibinfo{pages}{443--458}.
\bibitem[{Mincer(1974)}]{mincer74}
\bibinfo{author}{Mincer, J.}, \bibinfo{year}{1974}.
\newblock \bibinfo{title}{Schooling, Experience, and Earnings}.
\newblock \bibinfo{publisher}{National Bureau of Economic Research (NBER)}.
\bibitem[{Murphy and Welch(1990)}]{murphy90}
\bibinfo{author}{Murphy, K.}, \bibinfo{author}{Welch, F.},
  \bibinfo{year}{1990}.
\newblock \bibinfo{title}{Empirical age-earnings profiles}.
\newblock \bibinfo{journal}{Journal of Labor Economics} \bibinfo{volume}{8},
  \bibinfo{pages}{202--229}.
\bibitem[{Pashchenko(2013)}]{pashchenko2013accounting}
\bibinfo{author}{Pashchenko, S.}, \bibinfo{year}{2013}.
\newblock \bibinfo{title}{Accounting for non-annuitization}.
\newblock \bibinfo{journal}{Journal of Public Economics} \bibinfo{volume}{98},
  \bibinfo{pages}{53--67}.
\bibitem[{Peijnenburg et~al.(2016)Peijnenburg, Nijman and
  Werker}]{peijnenburg16}
\bibinfo{author}{Peijnenburg, K.}, \bibinfo{author}{Nijman, T.},
  \bibinfo{author}{Werker, B.}, \bibinfo{year}{2016}.
\newblock \bibinfo{title}{The annuity puzzle remains a puzzle}.
\newblock \bibinfo{journal}{Journal of Economic Dynamics and Control}
  \bibinfo{volume}{70}, \bibinfo{pages}{18--35}.
\bibitem[{Pliska and Ye(2007)}]{pliska07}
\bibinfo{author}{Pliska, S.}, \bibinfo{author}{Ye, J.}, \bibinfo{year}{2007}.
\newblock \bibinfo{title}{Optimal life insurance purchase and
  consumption/investment under uncertain lifetime}.
\newblock \bibinfo{journal}{Journal of Banking and Finance}
  \bibinfo{volume}{31}, \bibinfo{pages}{1307--1319}.
\bibitem[{Richard(1975)}]{richard75}
\bibinfo{author}{Richard, S.}, \bibinfo{year}{1975}.
\newblock \bibinfo{title}{Optimal consumption, portfolio and life insurance
  rules for an uncertain lived individual in a continuous time model}.
\newblock \bibinfo{journal}{Journal of Financial Economics}
  \bibinfo{volume}{2}, \bibinfo{pages}{187--203}.
\bibitem[{Shreve and Soner(1994)}]{shreve94}
\bibinfo{author}{Shreve, S.}, \bibinfo{author}{Soner, H.},
  \bibinfo{year}{1994}.
\newblock \bibinfo{title}{Optimal investment and consumption with transaction
  costs}.
\newblock \bibinfo{journal}{The Annals of Applied Probability}
  \bibinfo{volume}{4}, \bibinfo{pages}{609--692}.
\bibitem[{Yaari(1965)}]{yaari65}
\bibinfo{author}{Yaari, M.}, \bibinfo{year}{1965}.
\newblock \bibinfo{title}{Uncertain lifetime, life insurance, and the theory of
  the consumer}.
\newblock \bibinfo{journal}{The Review of Economic Studies}
  \bibinfo{volume}{32}, \bibinfo{pages}{137--150}.

\end{thebibliography}

\end{document}